\documentclass[twocolumn]{autart} 

\usepackage{amssymb}
\usepackage{amsmath}
\usepackage{amsfonts}                                
\usepackage{enumerate}
\usepackage{mathrsfs}
\usepackage{soul,xcolor}
\usepackage{tabulary}
\usepackage{cite}
\usepackage{graphicx}          
\usepackage{color}      
\usepackage{epsfig} % for postscript graphics files
\usepackage{epstopdf}
\usepackage{times} % assumes new font selection scheme installed
\def\qedp{\hspace*{\fill}~{\tiny $\blacksquare$}}
\def\qed{\relax\ifmmode\hskip2em \Box\else\unskip\nobreak\hskip1em $\Box$\fi}

\setstcolor{red}
\usepackage{soul}
\usepackage{enumitem}
\setlist{nolistsep}

\newlength{\bibitemsep}
\newlength{\bibparskip}
\let\oldthebibliography\thebibliography
\renewcommand\thebibliography[1]{%
	\oldthebibliography{#1}%
	\setlength{\parskip}{\bibitemsep}%
	\setlength{\itemsep}{\bibparskip}%
}

\def\qedp{\hspace*{\fill}~{\tiny $\blacksquare$}}
\def\qed{\relax\ifmmode\hskip2em \Box\else\unskip\nobreak\hskip1em $\Box$\fi}

\newtheorem{theorem}{Theorem}
\newtheorem{itlemma}{Lemma}
\newtheorem{itdefinition}{Definition}
\newtheorem{itproposition}{Proposition}
\newtheorem{itresult}{Result}
\newtheorem{itremark}{Remark}
\newtheorem{itassumption}{Assumption}
\newtheorem{itcorollary}{Corollary}
\newtheorem{itexample}{Example}

\newenvironment{proposition}{\begin{itproposition}\rm}{\end{itproposition}}
\newenvironment{remark}{\begin{itremark}\rm}{\end{itremark}}

\newenvironment{lemma}{\begin{itlemma}\rm}{\end{itlemma}}

\AtBeginDocument{
	\addtolength{\abovedisplayskip}{-1.5ex} %upper space of equation
	\addtolength{\abovedisplayshortskip}{-1.5ex}
	\addtolength{\belowdisplayskip}{-1.5ex}%lower space of equation
	\addtolength{\belowdisplayshortskip}{-1.5ex}
}

\begin{document}

	\begin{frontmatter}

		\title{Asymptotic tracking control of dynamic reference over homomorphically encrypted data with finite modulus \thanksref{footnoteinfo} } % Title, preferably not more 
		% than 10 words.
		\vspace{-3mm}
		\thanks[footnoteinfo]{Corresponding author: Junsoo Kim. The material in this paper was not presented at any conference. }
		
		\vspace{-7mm}

		\author[First]{Shuai Feng}\ead{s.feng@njust.edu.cn},    % Add the 
		\author[Second]{Junsoo Kim}\ead{ junsookim@seoultech.ac.kr}               % e-mail address 
		%		\author[First]{2}\ead{2},  
		%		\author[Third]{3}\ead{3},  % (ead) as shown
		%		\author[First]{4}\ead{4}
		\vspace{-1mm}
		\address[First]{School of Automation, Nanjing University of Science and Technology, Nanjing 210094, China}  				\vspace{-1mm}                                             
		\address[Second]{Department of Electrical and Information Engineering, Seoul National University of Science and Technology, Korea}            
		%		\vspace{-1mm}
		%		\address[Third]{1}      

		% help of the Automatica 
		% keyword wizard

		\begin{abstract}                          % Abstract of not more than 200 words.
			This paper considers a tracking control problem, in which the dynamic controller is encrypted with an additively homomorphic encryption scheme and the output of a process tracks a dynamic reference asymptotically. Our paper is motivated by the following  problem: When dealing with both asymptotic tracking and dynamic reference, we find that the control input is generally subject to overflow issues under a finite modulus, though the dynamic controller consists of only integer coefficients. First, we provide a new controller design method such that the coefficients of the tracking controller can be transformed into integers leveraging the zooming-in factor of dynamic quantization. 
			 By the Cayley-Hamilton theorem, we represent the control input as linear combination of the previous control inputs. Leveraging the property above, we design an algorithm on the actuator side such that it can restore the control input from the lower bits under a finite modulus. A lower bound of the modulus is also provided. 
			As an extension of the first result, we further solve the problem of unbounded internal state taking place in the actuator. In particular, the actuator can restore the correct control input under the same modulus. 
			A simulation example is provided to verify the control schemes proposed in our paper.
			%		\vspace{-2mm}	 
		\end{abstract}
		%		\vspace{-8mm}
	\end{frontmatter}

	%%%%%%%%%%%%%%%%%%%%%%%%%%%%%%%%%%%%%%%%%%%%%%%%%%%%%%%%%%%%%%%%%%%%%%%%%%%%%%%%
	\section{Introduction}
	
	The recent decade has witnessed revolutions in communication and computation technologies. Leveraging 5G/Wifi 6, edge and cloud computation to name a few, cyber physical systems (CPSs) can improve the performance and meanwhile even reduce the budget. In general, cloud computing service is outsourced to a third party. Therefore, the challenges of protecting the confidentiality of the CPSs’ data operated in the cloud computing centers arise \cite{kim2022comparison, schluter2023brief, darup2021encrypted}.

	%Deceptive attacks influence the integrity of data and DoS attacks can induce malicious packet drops by radio-frequency interference and/or flooding the target with an overwhelming flux of packets to name a few. 

	Homomorphic encryption (HE) is promising in simultaneously allowing for data operation and securing its confidentiality during cloud computing processes. To be specific, the computation can be directly performed over encrypted data without decryption. After the initial attempt of applying HE to control systems\cite{kogiso2015cyber}, various HE-based algorithms have been established to deal with networked control\cite{farokhi2017secure}, encrypted MPC\cite{darup2017towards}, secured consensus\cite{ruan2019secure}, cloud-based optimization\cite{alexandru2020cloud}, state estimation\cite{zhang2020secure} and formation control\cite{marcantoni2022secure} to name a few.

	The overflow problem, in which the controller's state/control input exceeds the modulus of a cryptosystem (the size of the message space), is one of the most critical problems when applying HE to dynamic control systems. Therefore, the overflow issue in HE-based control has attracted substantial attention, and the representative solutions such as re-encryption\cite{kogiso2015cyber}, state reset\cite{murguia2020secure} and controller having only integer coefficients\cite{cheon2018need,schluter2021stability} have been proposed. By exploiting observability and re-encryption, the recent paper\cite{kimtac1} proposes a framework such that any dynamic controller can be transformed into a comparable form consisting of only integer coefficients. The comparable dynamic controller can prevent overflow and operate for infinite horizon without resetting and bootstrapping.
	%	 Shortly, we will show that even though a controller has only integer coefficients, overflow issue can still occur  when one deals with asymptotic tracking control of a dynamic reference.
	%	

	%Different from the practical stabilization results in \cite{kimtac1,kim2024further}, the papers \cite{kimcdc, teranishi2019stability} integrate dynamic quantization with HE to achieve asymptotic stabilization.    

	In \cite{kimcdc}, the authors propose a HE-based control scheme to realize asymptotic stabilization given a constant reference signal. 
	When applying the algorithm in \cite{kimcdc} to deal with tracking control, we find the following issues (see Section 2.3):
	\begin{enumerate}[topsep=-10pt, partopsep=-10pt] 
		\item[1)] Overflow will occur in general, under a finite modulus, when one deals with \emph{asymptotic} tracking and a \emph{dynamic} reference.
		\item[2)] Overflow does not occur under a finite modulus, when one aims at achieving practical tracking of a dynamic reference.  
		\item[3)] Overflow does not occur under a finite modulus, when one deals with asymptotic tracking of a constant reference.
	\end{enumerate} 	
	%	Since the plaintext space of HE supports only integers, quantization of signals is necessary.
	%the non-integer state is quantized into integers for encryption}. 
The result in \cite{cheon2018need} suggests that a controller with integer coefficients can solve the overflow problem caused by the multiplication between controller's state and non-integer numbers. However, even when the ``original controller" has integer coefficients as suggested in \cite{cheon2018need}, the overflow problem in 1) cannot be fixed. 
This is because the control input is also amplified by the scaling factor in the dynamic quantization mechanism. 
If overflow occurs, it will be difficult for the actuator to restore the correct control input. In an ideal case, if the cryptosystem is allowed to have an unlimited modulus, i.e., the size of the message space is infinitely large, there is no overflow issue. However, in practice, the modulus is limited.

%Then, one converts the controller into a comparable form having integer coefficients only \cite{cheon2018need,kimtac1}. 
%%We find that overflow can still occur, even though the controller has only integer coefficients. 
%%This is because as the scaling factor in dynamic quantization approaches zero, the control input is amplified to infinity.
% Note that overflow does not occur in practical stabilization control \cite{kimtac1,kim2024further}, and can be well tackled in asymptotic stabilizing control, i.e., the reference signal is static \cite{kimcdc}. As discovered in our paper, if one simultaneously deals with \textbf{asymptotic} tracking and \textbf{dynamic} reference, overflow can occur

%Therefore, to realize the control objectives in the paper, it is important that the actuator is able to restore the correct control input. 

%, which makes control problems under limited modulus challenging. 
%Thus, our control schemes do not suppose infinity large modulus. In particular, we provide a lower bound of the modulus for each control scheme.  

This paper aims to solve a tracking control problem by HE with a finite modulus, in which the output of a process should track a dynamic reference, asymptotically.
%We design a dynamic controller with only integer coefficients by exploiting dynamic quantization.
First, we provide a conventional controller to realize quantized asymptotic tracking control by the results in output regulation/tracking \cite{Francis, aguiar2008performance} and dynamic quantization\cite{brockett2000quantized,feng2020tac,teranishi2019stability,kimcdc}, in which the controller coefficients are not necessarily integers. 
To make the controller have all integer coefficients, 
instead of using matrix conversion\cite{kimtac1}, 
we use the zooming-in factor in dynamic quantization to scale controller parameters into integers. By such a method, re-encryption that was utilized in \cite{kogiso2015cyber, kimtac1} is not required in this work. 
Leveraging the Cayley–Hamilton theorem\cite{horn2012matrix}, we represent the control input as a linear combination of previous control inputs. By doing so, on the actuator side, we design an algorithm such that the actuator can restore the control input from the lower bits, taking advantage of the previous control inputs stored in the actuator's memory. 
Importantly, this work provides an explicit lower bound of the modulus.
	Although a finite modulus is chosen and unchanged with time, it will be shown that asymptotic tracking of a dynamic reference can be achieved without the overflow problem.

By extending the first result, in the second part of the paper, we briefly address the problem of unbounded internal state that exists in \cite{kimcdc} and also in the first control scheme of our paper. It refers to the following issue: the actuator needs to generate an internal integer state, whose norm will grow to the infinity.  
%	However, in the field of control, one in general should design a control system whose state is bounded. Moreover, in practice, numerical saturation in electronic devices exists and impacts the unbounded internal state. That is, if numerical saturation occurs, actuator fails to generate the internal state and hence fails to restore correct the control input. 
	To solve this problem, instead of directly transmitting the cipher control input as in the first control scheme, the controller transmits a linear combination of current and previous cipher control inputs. In particular, for restoring the correct control input, the cryptosystem requires the same modulus as that in the first control scheme.

This paper is organized as follows. In Section 2, we introduce the overflow issue in asymptotic tracking of a dynamic reference under HE,  and the control objectives. Section 3 presents the main result and its extension solving the problem of unbounded internal state. 
A numerical example is presented in Section 4, and finally Section 5 ends the paper with conclusions.

\textbf{Notation.} We let $\mathbb R$, $\mathbb{Q}$ and $\mathbb Z$ denote the sets of real, rational and integer numbers, respectively. 
For $b\in \mathbb R$, let $\lfloor b \rfloor$ be the floor function such that $\lfloor b \rfloor= \max\{c\in \mathbb{Z}\,|\,c\le b\}$. 
We let $I_n$ denote the identity matrix with dimension $n$ and $\mathbf 0:=[0\,\,0\cdots0]^T$.
Let $\chi \in \mathbb{R}$ be a scalar before quantization and $q_{t}(\cdot)$ be the quantization function such that
$
q_{t} (\chi)= 	\psi$ if $(2\psi-1)/2 \le \chi  <  (2\psi+1)/2$, and $q_{t} (\chi)= -q_{t} (-\chi)$ if $ \chi \le - \frac{1}{2}$, 	
where $\psi =0, 1, 2, \cdots$. 
%If the quantizer is unsaturated such that $|\chi| \le (2R+1)\sigma $, then the  quantization error satisfies 
%$
%|\chi - q_{p}(\chi)| \le \sigma.
%$
The vector version of the quantization function is defined as $Q(x):= [\,q_t(x_1)\,\,q_t(x_2)  \cdots  q_t(x_n) \, ]^T \in \mathbb Z ^ n$, where $x = [x_1\,\, x_ 2  \cdots x_n]^ T \in \mathbb R ^n$.   
%In the subsequent sections, by quantizer overflow or saturation, we mean $\chi$ exceeds the range of quantization, i.e., $|\chi|> (2R+1)\sigma$. 
For a vector $x$ and a matrix $\Omega$, let $\|x\|$ and $\|x\|_\infty$ denote the $2 $- and $\infty$-norms of $x$, respectively,  
and $\|\Omega\|$ and $\|\Omega\|_\infty$ represent the corresponding induced norms of matrix $\Omega$. Moreover, $\rho(\Omega)$ denotes the spectral radius of $\Omega$. 
Let $\mathbf 0$ denote a vector with only zero elements in a compatible size. 
For $g\in \mathbb Z$ and $q\in \{1, 2, \cdots\}$, the modulo operation is defined by $g\,\,\text{mod}\,\,q:=g- \lfloor \frac{g}{q}\rfloor q$.
The set of integers modulo $q\in \mathbb Z_{\ge 1}$ is denoted by $\mathbb Z_q =\{0, 1, \cdots, q-1\}$. For a vector $v=[v_1\,\, v_2  \cdots  v_n] \in  \mathbb Z^n$, $v\,\text{mod}\, q$ denotes the element-wise modulo operation such that $v\,\,\text{mod}\,\, q= [v_1\,\text{mod}\,q\,\,\, v_2\,\text{mod}\,q \cdots  v_n\,\text{mod}\,q]^T$. 
%In the remainder of the paper, we will not distinguish ``\text{mod}" and ``$\text{mod}_v$" for simplicity. 
In this paper, by an ``\textbf{integer matrix}" and an ``\textbf{integer vector}", they refer to a matrix and a vector, respectively, whose elements are all integers. Any integer vector $v$ can be written as $v= v_1 q + v \,\text{mod}\,q$ in which $v_1$ is some integer vector having the same dimension of $v$ and $q\in \mathbb{Z}$. In our paper, we abuse the notations ``higher bits" and ``lower bits" such that the ``higher bits" refers to $v_1$ and the ``lower bits" refers to $v \,\text{mod}\,q$.

\section{Problem formulation}

\subsection{Preliminaries of additively homomorphic encryption}

%For some $q\in \mathbb Z_{\ge 1}$, $\mathbb Z_q$ is the space of plaintexts (for a scalar). Let $\mathcal C$ denote the space of ciphertexts. Let $\mathbf{Enc(\cdot)}: \mathbb Z_q \to \mathcal C$ and $\mathbf{Dec(\cdot)}: \mathcal C\to \mathbb Z_q $ denote the encryption and decryption processes, respectively. Furthermore,

There are two types of partially homomorphic encryption schemes, additively and multiplicatively homomorphic encryptions. For example, the Paillier and the ElGamal cryptosystems are well-known additively and multiplicatively homomorphic encryption algorithms, respectively \cite{Paillier,elgamal1985public}.   

In our paper, we apply additively homomorphic encryption to secure the control systems. 
In an additively homomorphic encryption scheme, the spaces of plaintexts and ciphertexts are $\mathbb Z_q ^n$ and $\mathcal C^n$, respectively. For a vector in $\mathbb Z_q ^n $, its encryption and decryption processes are given by $\mathbf{Enc}(\cdot): \mathbb Z_q ^n \to \mathcal C^n$ and $\mathbf{Dec(\cdot)}: \mathcal C^n \to \mathbb Z_q ^n$, respectively.
%In this paper, the cryptosystem is considered to be additively homomorphic. 
Secret and public keys are also involved in $\mathbf{Enc}(\cdot)$ and $\mathbf{Dec(\cdot)}$, but for simplicity they are omitted. 
The properties of additive homomorphic encryption are listed as follows:
\begin{enumerate}[topsep=-10pt, partopsep=-10pt] 
	\item[1.] For $x \in  \mathbb Z_q ^n$, one has $\mathbf{Dec}(\mathbf{Enc}(x))= x$.
	\item[2.] Consider the ciphertexts $\mathbf c_1\in \mathcal  C ^n$ and $\mathbf c_2 \in \mathcal  C ^n$.
	%, and their ciphertexts $\mathbf{Enc}(v_1) \in \mathcal C^n$ and $\mathbf{Enc}(v_2) \in \mathcal C^n$, respectively. 
	There exists an operation $\oplus  $ such that $\mathbf{Dec}(\mathbf c_1 \oplus \mathbf c_2)= \mathbf{Dec}(\mathbf c_1)+ \mathbf{Dec}(\mathbf c_2) \,\, \text{mod}\,\,q$. 
	\item[3.]  Consider a matrix in plaintext $M \in \mathbb Z_q ^{m \times n}$ and a vector in ciphertext $\mathbf c_3 \in  \mathcal C ^n$. There exists an operation ``$\cdot$" representing multiplication such that $\mathbf{Dec}(M \cdot \mathbf c_3)= M \mathbf{Dec}(\mathbf c_3) \,\, \text{mod}\,\,q$. 
\end{enumerate}
%$\bullet$ \\
%$\bullet$ \\
%$\bullet$ 

In light of the preliminaries above, we call $q$ the modulus of a cryptosystem, which should be a finite integer to be shown later. For more information about additively homomorphic encryption algorithms and their properties, we refer the readers to the survey papers \cite{kim2022comparison, schluter2023brief, darup2021encrypted} and \cite{Paillier}. 

\subsection{Tracking control of a dynamic reference}
In this paper, we consider a discrete-time process
\begin{subequations}\label{process}
	\begin{align}
		x_p(k+1)&=Ax_p(k)+Bu(k)\\
		y_p(k)&=Cx_p(k)
	\end{align}
\end{subequations}
where $x_p(k)\in \mathbb{R}^{n}$ denotes the state of the process, $u(k) \in \mathbb{R}^{w}$ denotes the control input and $y_p(k)\in \mathbb{R}^{v}$ denotes the output. We assume $A \in \mathbb Q^{n \times n}$, $B \in \mathbb Q^{n \times w}$, $C\in \mathbb Q^{v\times n}$,
% because any irrational number can be approximated by a rational number with infinite precision \cite{kimtac1}. 
$(A,B)$ stabilizable and $(A,C)$ observable. Namely, there exist $K \in \mathbb Q ^{w \times n}$ and $L \in \mathbb Q ^{n\times v} $ such that $\rho(A+B K)<1$ and $\rho(A-LC)<1$, respectively. We also assume that the initial condition $x_p(0)$ is not ``infinitely large" such that there exists $C_{x_p(0)}$ satisfying $\| x_p(0)\|_\infty \le C_{x_p(0)}$\cite{1186780,feng2024bottleneck}. 
%Note that $C_{x_p(0)} $ can be arbitrarily large as long as it satisfies this bound.
%\textcolor{red}{Hi Junsoo: I am afraid that we have to assume the matrices in (1) should have rational-number elements. If we assume they are real matrices and approximate them with rational numbers, we need an additional strong assumption to achieve asymp-stability. I will show you in our meeting.}
%{\color{orange}(Thanks, this may be related to use of re-encryption, or use of $1/\gamma$ for the implementation)}

This paper aims at solving a tracking control problem. The dynamics of the reference is described by an exosystem
\begin{align}\label{reference}
	v_p(k+1) = S v_p(k)  
\end{align}
with $v_p(k) \in \mathbb R ^{v}$ and $S \in \mathbb Q^{v\times v}$. 
Similarly, we assume that there exists $C_{v_p(0)} $ such that $\| v_p(0)\|_\infty \le C_{v_p(0)} $ \cite{1186780,feng2024bottleneck}. 
We say that the process can asymptotically track the reference if
\begin{align}\label{tracking}
	\lim_{k  \to \infty } \|  y_p(k)- v_p(k)\|_\infty = 0. 
\end{align}
To make the problem meaningful, we assume $\rho(S) \ge 1$. Otherwise, one can simply design a stabilizing controller to steer $x_p(k)\to 0$ without involving the reference $v_p(k)$. 

%If $\rho(S)<1$, one should have $v_p(k)\to 0$ as $k\to \infty$. Then, any feedback controller that stabilizes (\ref{process}) is sufficient to solve the tracking control problem. 

\subsection{Overflow issue in asymptotic tracking control}
We assume that there exists a pair of matrices $(\Gamma  \in \mathbb Q^{n \times v}, V \in \mathbb Q ^{w \times v})$ satisfying
$
\Gamma S=A \Gamma  + B V$ and $ I_v	= C \Gamma 
$, where $I_v$ is the identity matrix with dimension $v$. This is a standard assumption in regulation/tracking problems. We refer the readers to the seminal paper \cite{Francis} and its citations for more information. 

The problem (\ref{tracking}) has been well studied (e.g., in \cite{Francis, aguiar2008performance}). It can be solved by the following controller
\begin{subequations}\label{controller 1}
	\begin{align}
		\hat x(k+1) &= A \hat x(k) + B u(k) + L \left(y_p(k) -C \hat x(k)\right) \\
		u(k) &= K \hat x(k) + (V-K \Gamma)  v_p(k)   
	\end{align}
\end{subequations}
in which $\rho(A+B K)<1$, $\rho(A-LC)<1$, $\hat x(k) \in \mathbb R ^n$ is the state of the observer (\ref{controller 1}a). To apply homomorphic encryption schemes, $y_p(k)$ and $v_p(k)$ should be received in the form of integer vectors. This requires quantization. Thus, (\ref{controller 1}) involving quantized information can be described by
\begin{subequations}\label{controller 2}
	\begin{align}
		\hat x(k+1) &= A \hat x(k) + B u(k)  \nonumber\\
		& \quad + L \left(l(k)Q\left(\frac{y_p(k)}{l(k)}\right) -C \hat x(k)\right) \\
		\hat v(k+1) &= S \hat v(k) + l(k) Q\left( \frac{S v_p(k) - S \hat v(k)}{l(k)}\right)\\
		u(k) &= K \hat x(k) + (V-K \Gamma) \hat v(k)  
	\end{align}
\end{subequations}
in which $\hat v(k)$ is the estimation of $v_p(k)$ and $Q(\cdot)$ is the quantization function in the Notation. In (\ref{controller 2}), the scaling factor updates as follows
\begin{align}
	l(k+1) = \gamma l(k), \quad 0< \gamma <1,\quad l(0) \in \mathbb{R}_{>0}
\end{align}
where $\gamma$ is the so-called zooming-in factor in dynamic quantization\cite{brockett2000quantized}. 
%Note that, the purpose of dynamic quantization, i.e., $\lim_{k\to \infty} l(k) = 0$, is achieving asymptotic tracking control (\ref{tracking}). Otherwise, $l(k)$ can be a constant or lower bounded, under which the tracking error will not converge to 0.   
By defining 
\begin{align}
	&\bar x(k) :=\! \frac{\hat x(k)}{l(k)}, \bar y(k):=\! \frac{y_p(k)}{l(k)}, \bar u(k) :=\!  \frac{  u(k)}{l(k)}, \nonumber\\
	&\bar v_p(k): = \frac{v_p(k)}{l(k)}, \bar v(k):=\!   \frac{  \hat v(k)}{l(k)} \nonumber
\end{align}
the controller (\ref{controller 2}) can be written as
\begin{subequations}\label{controller 3}
	\begin{align}
		\bar x(k+1) &= \frac{A-LC}{\gamma} \bar x(t) + \frac{B}{\gamma} \bar u(t) + \frac{L}{\gamma}   Q\left(\bar y(k)   \right) \\
		\bar v(k+1) &= \frac{S}{\gamma} \bar v(k) + \frac{1}{\gamma}  Q ( S \bar v_p(k) - S \bar  v(k)   ) \\
		\bar	u(k) &= K \bar x(t) + (V-K \Gamma) \bar v(k).  
	\end{align}
\end{subequations}
%It is simple to see that $l(k)\to 0$ as $k \to \infty$. It is an important parameter because if $l(k)\to \epsilon >0$, asymptotic tracking control cannot be guaranteed. 
By the results in \cite{cheon2018need}, one needs to transform the matrices in (\ref{controller 3}), particularly $(A-LC)/\gamma$ and $S/\gamma$, to integer matrices.   
In this section, for the ease of presenting the overflow issue in encrypted tracking control, we assume that $A, B, L, K$ and $V-K \Gamma$ in (\ref{controller 1}) are integer matrices, under which the dynamic controller should have been free of overflow\cite{cheon2018need}. Moreover, we assume that $\gamma$ is selected such that $(A-LC)/\gamma$, $S/\gamma$, $B/\gamma$, $L/\gamma$, $S/\gamma$, $1/\gamma$ in (\ref{controller 3}) consist of integers. The second assumption above will be removed by the results later in Lemma \ref{lemma 1}.

%It is known that an encrypted controller cannot run for an infinite horizon if a dynamic controller consists of non-integer parameters. 
\textbf{Constant reference:} For a constant reference, i.e., $S$ in (\ref{reference}) is an identity matrix, we present the algorithm of restoring the control input $u(k)$ on the actuator side\cite{kimcdc}. Let $\mathbf{\bar u}(k)$ denote the ciphertext of $\bar u(k)$. The algorithm on the actuator side in is given as      
\begin{align}\label{9}
	u_a(k)\!&=\! l(k)\!\! \left(\!\! \mathbf{Dec}(\mathbf { \bar  u }(k)) \!- \! \Big\lfloor \frac{\mathbf{Dec}(\mathbf { \bar  u }(k)) - \frac{  u(k-1)}{ l(k)}   + \frac{ q}{2}}{q} \Big\rfloor  q \!\right) \nonumber\\
	%&= l(t)\left( \frac{u(t)}{ l (t)}  - \Big\lfloor \frac{ \frac{u(t)}{l(t)}- \frac{  u(t-1)}{ l(t)}   + q/2}{q} \Big\rfloor  q\right) \nonumber\\
	\!&\! = l(k) \left(\! \frac{u(k)}{ l (k)} \! -\! \Big\lfloor \frac{ \frac{u(k)-u(k-1)}{l(k)}  + \frac{ q}{2}}{q} \Big\rfloor  q\! \right)\!=\!u(k).
\end{align}
%For the second equation in (\ref{9}), we refer the readers to Lemma 2 in \cite{kimcdc}.
Note that the last equality in (\ref{9}) holds because there must exist a finite $q$ such that $\| \frac{u(k)-u(k-1)}{l(k)}  \|_\infty < \frac{q}{2}$ for all $k$. For more details about (\ref{9}), we refer the readers to Lemma 2 in \cite{kimcdc}.

%In case $v_p(k)$ in (\ref{controller 1}) is a static reference, by substituting (\ref{controller 1}c), one can verify that 
%\begin{align}\label{10}
%	&\frac{u(k)-u(k-1)}{l(k)} \nonumber\\
%	&= (K \hat x(k) + (V-K \Gamma) \hat v(k)   \nonumber\\
%	& \quad- (K \hat x(k-1) + (V-K \Gamma) \hat v(k-1))  ) / l(k) \nonumber\\
%	&= K (\hat x(k)- \hat  x(k-1))  / l(k) \nonumber\\
%	&= K(x(k)+ e(k) - x(k-1)- e(k-1)) / l(k) \nonumber\\
%	&= K(\delta_x(k)- \delta_x(k-1)+ e(k) - e(k-1)) / l(k)
%\end{align}
%in which $e(k) := \hat x(k) - x(k)$ denotes the observer estimation error and $\delta_x(k): = x(k)-x_\infty$ denotes the discrepancy between $x(k)$ and its equilibrium $x_\infty$. By the last equation of (\ref{10}), as long as $l(k)$ converges slower than $\delta_x(k)$ and $e(k)$, $\frac{u(k)-u(k-1)}{l(k)}$ has an upper bound and therefore there exists a finite $q$ such that $u(k)$ can be constructed by the actuator with (\ref{9})\cite{kimcdc}. 

\textbf{Dynamic reference:} In the following, we show that when the process tracks the dynamic reference (\ref{reference}) (i.e., $S$ is not an identity matrix and $\rho(S)\ge 1$), the actuator cannot restore $u(k)$ by (\ref{9}). 
Specifically, substituting (\ref{controller 2}c), one has
\begin{align}\label{11}
	&\frac{u(k)-u(k-1)}{l(k)}  \nonumber\\
	&= (K \hat x(k) + (V-K \Gamma) \hat v(k)   \\
	&\quad- (K \hat x(k-1) + (V-K \Gamma) \hat v(k-1)) ) / l(k) \nonumber \\
	&=( Kp(k) \!-\! K p(k-1)  \!+\!  V\hat v(k) \!-\! V \hat v(k-1)) / l(k) \to \infty \nonumber
\end{align}
in which $p(k)/l(k)$ and $p(k-1)/l(k)=p(k-1)/(\gamma l(k-1))$ are upper bounded to be shown in the proof of Theorem \ref{Theorem 1} but $\hat v(k)-  \hat v(k-1)$ does not converge to zero. Thus, as $l(k) \to 0$, one must have $\frac{u(k)-u(k-1)}{l(k)} \to \infty$. This implies that for any bounded $q$, one will surely encounter $\|\frac{u(k)-u(k-1)}{l(k)}\|_\infty > \frac{q}{2}$, i.e. \textbf{overflow issue}, and it will be impossible to restore $u(k)$ from $\mathbf{Dec}(\mathbf { \bar  u }(k))$ by (\ref{9}) when overflow occurs. Note that if $l(k)$ is lower bounded, one can avoid the overflow issue. However, it is not possible to realize asymptotic tracking control (\ref{tracking}). One can only achieve practical tracking, namely, $\|y_p(k) - v_p(k)\|_\infty < \epsilon$ for some $\epsilon >0$. Please note that $\lim_{k\to\infty}\hat v(k)-  \hat v(k-1)\ne 0$ in (\ref{11}) is not due to the application of $\hat v(k)$ in $u(k)$. Assume that (\ref{controller 2}b) is perfectly designed such that $\hat v(k)=v_p(k)$ for all $k$. In light of (\ref{11}), one still has $\frac{u(k)-u(k-1)}{l(k)}\to \infty$. At last, we emphasize that a controller with integer coefficients cannot solve the overflow problem because we have derived (\ref{11}) by assuming that $A, B, L, K$ and $V-K \Gamma$ in (\ref{controller 1}) are integer matrices.

In a nutshell, in encrypted control problems, when dealing with dynamic reference and asymptotic tracking simultaneously, one would encounter overflow issues under a finite modulus, though the dynamic controller consists of only integer matrices. 
%Thus, $\|\bar u(k)\|_\infty$ becomes infinitely larger as $k \to \infty$. 
%At some $k$, $\frac{u(k)-u(k-1)}{l(k)} > \frac{q}{2}$ occurs and hence it is not possible to obtain the exact $u(k)$ by (\ref{9}). 
%In the following, we present the reasons of overflow for the case of time-varying reference tracking. 
%It is simple to see that overflow of $\bar u(k)$ occurs because of the time-varying reference $\bar v(k) \to \infty$ as $k \to \infty$. 

\textbf{Control objectives:}
In view of the process (\ref{process}) and dynamic reference (\ref{reference}),
\begin{itemize}[topsep=-10pt, partopsep=-10pt] 
	\item[1.]  design controllers operated over encrypted data utilizing additively homomorphic encryption;
	\item[2.]  design algorithms on the actuator side that can restore the control input $u(k)$ in (\ref{controller 2}c) from encrypted messages under a finite $q$
\end{itemize}
such that asymptotic tracking control (\ref{tracking}) is realized.  

The control schemes to be designed should be subject to the constraints 1)-3) in Section II-B in \cite{kimtac1}. Moreover, in our paper, we present two additional constraints.
i) The actuator does not have access to the reference $v_p(k)$ and the process output $y_p(k)$.  
ii) The actuator does not perform ``re-encryption". %The controller does not have access to the secret key and hence cannot decrypt the ciphertexts. 
%The reasons of imposing i) and ii) will be specified later. 

%The constraint above is to prevent 
%Compared with those constraints in \cite{kimtac1}, 

%To make the parameters (matrices) in (\ref{controller 3}) have all integers, one can use re-encryption. However, re-encryption is computationally intensive. The integer parameter transition can be solved if there exists a sufficiently small $\gamma$. 

\section{Encrypted tracking control}

\subsection{Encrypted controller design and finite modulus}

We first transform (\ref{controller 3}) into a controller with integer coefficients.
We define the following state
\begin{align}\label{12}
	\tilde x(k) := s \bar x(k),\,\,  \tilde v_p(k): = s \bar v_p(k),\,\,	\tilde v(k) := s \bar v(k)
\end{align}
with $0<s<1$.
Then, (\ref{controller 3}) can be transformed into 
\begin{subequations}\label{controller 4}
	\begin{align}
		\tilde  x(k+1)\! &=\! \frac{A-LC}{\gamma} \tilde x(k) \!+\! \frac{s B}{ \gamma} \bar u(k) \!+\! \frac{s L}{\gamma}   Q\!\left( \bar y(k)    \right) \\
		\tilde v(k+1)\!&=\! \frac{S}{\gamma} \tilde v(k) \!+\! \frac{s}{\gamma}  Q\left( \frac{S}{s} \tilde v_p(k) - \frac{S}{s} \tilde v(k) \right)\\
		\bar	u(k) \! &=\! \frac{K}{s} \tilde x(k) +  \frac{V-K \Gamma}{s} \tilde v(k) 
	\end{align}
\end{subequations}
with $\tilde x(0)= s\hat x(0)/l(0)$ and $\tilde v(0)= s \hat v(0)/l(0)$.

\begin{lemma}\label{lemma 1}
	Consider the controller (\ref{controller 4}). Its matrices can be converted to integer matrices by the following steps:
	\begin{itemize}[topsep=-10pt, partopsep=-10pt] 
		\item[1.] Select $(K, L)$ such that $\max\{\rho(A+BK), \rho( A-LC)\}$ is sufficiently small.
		\item[2.] 	Select $s \in \mathbb{Q}$ such that $\frac{S}{s} \in \mathbb{Z}^{v\times v}$ and $\frac{K}{s} \in \mathbb{Z}^{w\times n}$.
		\item[3.]  Choose $\gamma \in (\max\{\rho(A+BK), \rho(A-LC)\}, 1)$ such that 
		$\frac{A-LC}{\gamma} \in \mathbb{Z}^{n\times n} $, $\frac{s  B}{\gamma} \in \mathbb{Z}^{n\times w}$, $\frac{s L}{\gamma} \in \mathbb{Z}^{n\times v}$, $\frac{S}{\gamma}\in \mathbb{Z}^{v\times v}$ and $\frac{s }{\gamma} \in \mathbb Z$. Note that such a $\gamma$ always exists. 
	\end{itemize}
\end{lemma}

\textbf{Proof.}
%	There always exists $\gamma \in (\max\{\rho(A+BK), \rho(A-LC)\}, 1)$ such that $\frac{A-LC}{\gamma} \in \mathbb{Z}^{n\times n} $, $\frac{s  B}{\gamma} \in \mathbb{Z}^{n\times w}$, $\frac{s L}{\gamma} \in \mathbb{Z}^{n\times v}$, $\frac{S}{\gamma}\in \mathbb{Z}^{v\times v}$ and $\frac{s }{\gamma} \in \mathbb Z$. 
%	%To make the matrices above having only integer elements, 
To show the existence of $\gamma$ that scales the matrices in (\ref{controller 4}a)-(\ref{controller 4}b) into integer matrices, it is sufficient to show that $\gamma$ can be chosen arbitrarily small. One can always select $(K,L)$ such that $\max\{\rho(A+BK), \rho(A-LC)\}=0$.  Then, there must exist a sufficiently small $\gamma \in (0,1)$ such that the matrices in (\ref{controller 4}a)-(\ref{controller 4}b) are integer matrices. The reason of selecting $\gamma \in (\max\{\rho(A+BK), \rho(A-LC)\}, 1)$ is for ensuring closed-loop stability, which will be shown in the Appendix. For (\ref{controller 4}c), it is straightforward that there always exists a sufficiently small $s$ such that $\frac{S}{s} \in \mathbb{Z}^{v\times v}$ and $\frac{K}{s} \in \mathbb{Z}^{w\times n}$.   
\qedp

\begin{remark}
	By the results in Lemma \ref{lemma 1}, one can see that it is possible to scale the controller matrices into integer matrices by the zooming-in factor $\gamma \in (\max\{\rho(A+BK), \rho(A-LC)\}, 1)$. One may worry about that $\gamma$ is lower bounded by $\max\{\rho(A+BK), \rho(A-LC)\}$ and cannot be very small. However, since the eigenvalues of $A+BK$ and $A-LC$ can be arbitrarily placed, one can always place all their eigenvalues at 0 or arbitrarily close to 0. Hence, $\gamma$ is lower bounded by 0 or an arbitrarily small positive number, respectively. We emphasize that out method of converting controller matrices into integer matrices does not require re-encryption by observing that $\bar u(k)$ in (\ref{controller 4}a) can be taken from (\ref{controller 4}c), instead of being generated by the sensor or the actuator.  
	\qedp
	%Therefore, it is interesting to see that if the poles of the closed-loop dynamics can be arbitrarily placed, it is possible to render the controller matrices have all integer elements by leveraging dynamic quantization. Moreover, re-encryption is not necessary.  
\end{remark}

\begin{remark}
	The matrices in (\ref{controller 4}) can also be transformed into integer matrices by the approach in \cite{kimtac1} if $(S, V-K\Gamma)$ observable. First, 
	select $(K,L, L_0)$ such that $\max\{\rho(A+BK), \rho( A-LC), \rho(S- L_0(V-K\Gamma))\}=0$. Then,	we write (\ref{controller 4}b) into $\tilde v(k+1)= \frac{S}{\gamma} \tilde v(k) + \frac{s}{\gamma}  Q\left( \frac{S}{s} \tilde v_p(k) - \frac{S}{s} \tilde v(k) \right)- \frac{sL_0}{\gamma} \bar u(k) + \frac{sL_0}{\gamma}  \bar u(k) $. 
	Substituting $\bar u(k)$ in (\ref{controller 4}c) and $x_T(k): = T ^{-1} [\tilde x^T (k) \,\, \tilde v^T(k)]^T$ with $T=\text{diag}(T_1, T_2)$ invertible, (\ref{controller 4}) can be transformed into
	\begin{subequations}\label{reencryption}
		\begin{align}
			& x_T(k+1)= 
			\frac{1}{\gamma}
			T^{-1} 
			\begin{bmatrix}
				A\!-\!LC& \\
				L_0K &  S\!-\!L_0(V\!-\!K\Gamma) 
			\end{bmatrix}
			T x_T(k) \nonumber\\
			&+  
			\begin{bmatrix}
				\frac{sT_1 ^{-1}B}{\gamma} \\
				\frac{sT_2^{-1}L_0}{\gamma}
			\end{bmatrix}
			\bar u(k)+
			\begin{bmatrix}
				\frac{sT_1^{-1}L}{\gamma}Q(\bar y(k))\\
				\frac{sT_2^{-1}}{\gamma}Q(\frac{S}{s}\tilde v_p(k)- \frac{S}{s}\tilde v(k))\\
			\end{bmatrix}\\
			& \bar u(k)= \frac{1}{s}\left[KT_1\quad (V-K\Gamma)T_2\right] x_T(k).
		\end{align}
	\end{subequations}
	Because of $\max\{\rho( A-LC), \rho(S- L_0(V-K\Gamma))\}=0$, there must exist $T$ such that $ T^{-1} 
	[
	A\!-\!LC , 0 ;
	L_0K ,   S\!-\!L_0(V\!-\!K\Gamma) ]
	T$
	is an integer matrix. Afterwards, one can select sufficiently small $s$ and $\gamma$ such that $\frac{KT_1}{s}$, $\frac{(V-K\Gamma)T_2}{s}$, 
	$\frac{sT_1 ^{-1}B}{\gamma}$, 
	$\frac{sT_2^{-1}L_0}{\gamma}$, $\frac{s T_1 ^{-1}L}{\gamma}$ and $\frac{s T_2 ^{-1}}{\gamma}$ are integer matrices. 
	Note that re-encryption is still not required in (\ref{reencryption}). 
	Moreover, because the dynamic matrix of (\ref{reencryption}a) is a nilpotent matrix, (\ref{reencryption}) can be further transformed into the auto-regressive form $\bar u(k) = \sum_{i=1}^{n+v}K_1 \bar u(k-i) +\sum_{i=1}^{n+v} K_2 Q(\bar y(k-i)) +  \sum_{i=1}^{n+v}K_3  Q\left( \frac{S}{s} \tilde v_p(k-i) - \frac{S}{s} \tilde v(k-i) \right)$ for some integer matrices $K_1$, $K_2$ and $K_3$. For more information about the encrypted controller in the auto-regressive form, we refer the readers to \cite{teranishi2023input,lee2023encrypted}. \qedp
\end{remark}
%\textcolor{red}{Hi Junsoo: I've added Remark 2 to address that (12) can be transformed into integer controllers by the methods in your TAC regular paper. By the methods in your TAC regular paper, ARX model is also possible as a special case. However, to my impression, this remark may let the reviewer have too many questions... For example, how can one ensure the observability of $(S/\gamma, (V-K\Gamma)/s)$, or how to find $\gamma, s$ such that the pair is observable. I do not have an idea.... If you also think this is too risky, we then omit this remark...
	%}

%\begin{remark}
%The controller (\ref{controller 4}) can be converted to an auto-regressive form 
%\begin{align}
%u(k) = \sum_{i=1}^{n+v}a_i u(k-i) +\sum_{i=1}^{n+v} b_iy(k-i) 
%\end{align}
%\end{remark}

%In (\ref{controller 4}), $s \in \mathbb Q$ is chosen, possibly sufficiently small, such that $\frac{S}{s} \in \mathbb{Z}^{v\times v}$, $\frac{K}{s} \in \mathbb{Z}^{w\times n}$ and $ \frac{V-K \Gamma}{s} \in \mathbb{Z}^{w\times v}$. Moreover, $\gamma$ is a sufficiently small rational number such that $\frac{A-LC}{\gamma} \in \mathbb{Z}^{n\times n} $, $\frac{s  B}{\gamma} \in \mathbb{Z}^{n\times w}$, $\frac{s L}{\gamma} \in \mathbb{Z}^{n\times v}$, $\frac{S}{\gamma}\in \mathbb{Z}^{v\times v}$ and $\frac{s }{\gamma} \in \mathbb Z$. The existence of such a $\gamma$ will be specified in Remark \ref{remark 4}. 

By the results in Lemma \ref{lemma 1}, all the matrices in (\ref{controller 4}) contain only integer elements, and $\tilde x(0)$ and $\tilde v(0)$ are integer vectors under a sufficiently small $l(0)$.
%Then, we convert (\ref{controller 4}) into a form over $\mathbb Z_q$. 
By taking the modulo operation, we obtain the dynamics in (\ref{controller 4}) over $\mathbb Z_q$ 
%\begin{subequations}
%	\begin{align}
	%		&\tilde  x(k+1)   =\! \left[   \left(\frac{A-LC}{\gamma} \,\text{mod}\,q\right) \left(\tilde x(k) \,\text{mod}\,q\right)  \,\text{mod}\,q  \right. \nonumber\\
	%		& \quad+  \left(\frac{s B }{ \gamma} \,\text{mod}\,q\right) \left(\bar u(k) \,\text{mod}\,q\right) \,\text{mod}\,q  \nonumber\\
	%		& \quad \left. + \left(\frac{s L}{\gamma} \,\text{mod}\,q\right)  \left( Q\!\left( \bar y(k)    \right)  \,\text{mod}\,q \right)  \,\text{mod}\,q  \right] \,\text{mod}\,q \\ 
	%		&\tilde v(k+1)= \left[\! \left(\frac{S}{\gamma} \,\text{mod}\,q\right) \left(\tilde v(k) \,\text{mod}\,q \right) \,\text{mod}\,q \right.\nonumber\\
	%		&		+\left. \! \left(\frac{s}{\gamma}\,\text{mod}\,q\right) \left(\! Q\left(\! \frac{S}{s} \tilde v_p(k) - \frac{S}{s} \tilde v(k) \right) \,\text{mod}\,q \right) \right] \,\text{mod}\,q\\
	%		&\bar	u(k)  =\!\left[ \left(\frac{K}{s}\,\text{mod}\,q\right) \left(\tilde x(k)\,\text{mod}\,q\right) \,\text{mod}\,q \right. \nonumber\\
	%		& \left.+ \left(\frac{V-K \Gamma}{s}\,\text{mod}\,q\right) \left(\tilde v(k)\,\text{mod}\,q\right) \,\text{mod}\,q \right]  \,\text{mod}\,q .  
	%	\end{align}
%\end{subequations}
%It is straightforward that $\tilde  x(k+1)$, $\tilde v(k+1)$ and $\bar	u(k)$ also belong to the set $\mathbb Z_q$ because (\ref{controller mod}) is equivalent to 
\begin{subequations}\label{controller mod}
	\begin{align}
		\tilde  x(k+1)   &=\frac{A-LC}{\gamma} \tilde x(k) \! + \! \frac{s B }{ \gamma}\bar u(k)  \! +\! \frac{s L}{\gamma} Q\!\left( \bar y(k)    \right)   \,\text{mod}\,q  \\ 
		\tilde v(k+1)&\!= \!\frac{S}{\gamma}  \tilde v(k)  	\!+\! \frac{s}{\gamma}Q\left(\! \frac{S}{s} \tilde v_p(k) \!-\! \frac{S}{s} \tilde v(k) \! \right)  \,\text{mod}\,q\\
		\bar	u(k) 	& = \frac{K}{s} \tilde x(k) + \frac{V-K \Gamma}{s} \tilde v(k)  \,\text{mod}\,q  
	\end{align}
\end{subequations}
with initial conditions $\tilde x(0)$ mod $q$ and $\tilde v(0)$ mod $q$.
%\textcolor{red}{Hi Junsoo:  For (13), it is a detailed analysis, which aims at obtaining (14) and (15). I figured it out by myself and never discussed the correctness with you. If it is not correct, please help. Thanks! If this is correct, when space is limited, (13) can be omitted.}

\begin{figure}[t]
	\begin{center}
		\includegraphics[width=0.48\textwidth]{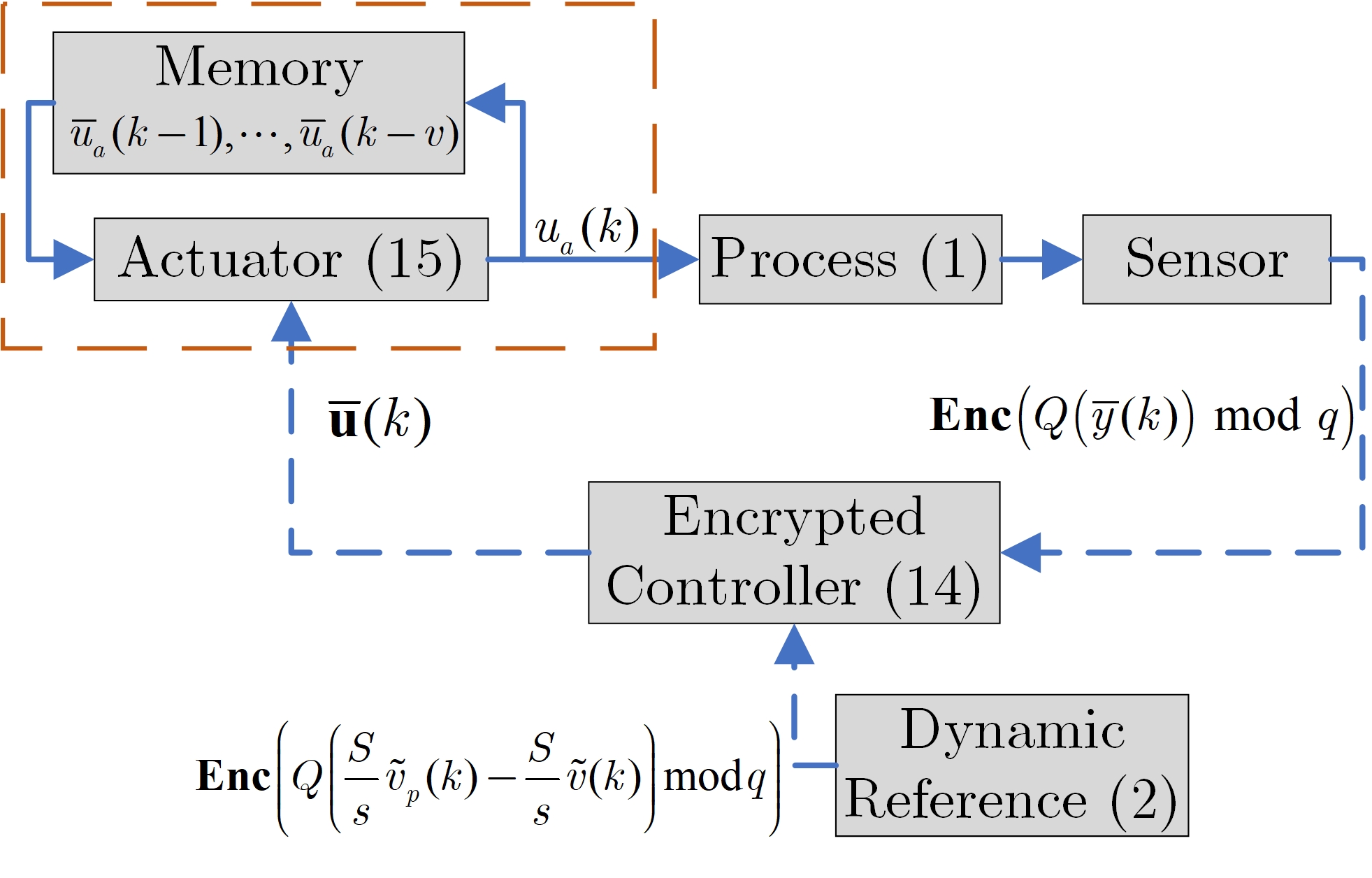}  \\
		\caption{Encrypted control architecture. Dashed lines represent networks. The sensor and reference provider transmit $\textbf{Enc}( Q(\bar y(k)   \,\,\text{mod}\,\,q  )$ and $\textbf{Enc}\left( Q\left( \frac{S}{s} \tilde v_p(k) - \frac{S}{s} \tilde v(k)\right)\text{mod}\,q  \right)$ to the encrypted controller over networks, respectively. The encrypted controller sends $\mathbf{\bar	u}(k) $ to the actuator over networks. The actuator computes $\bar u_a(k)$ based on $\bar u_a(k-j)$ ($j=1, \cdots, v$), and further computes $u_a(k)$. It feeds $u_a(k)$ to the process and stores $\bar u_a(k)$ in the memory for being utilized at $k+1$. } \label{Control archi 1}
	\end{center}
\end{figure}

In Fig. \ref{Control archi 1}, we present the control architecture over homomorphic encrypted data. The encrypted controller that computes cipher control inputs and the actuator that restores the control inputs from ciphertexts are designed as follows.

\textbf{Encrypted controller:} Based on (\ref{controller mod}) and the cryptosystem in Section 2.1, one can obtain the encrypted controller:
\begin{subequations}\label{controller 5} 
	\begin{align}
		\!\!\!\!\!	
		\mathbf	{\tilde  x} (k+1) \!&= (\mathbf{A-LC} ) \cdot	\mathbf	{\tilde  x}  (k) \oplus \mathbf{B} \cdot \mathbf  {\bar u}(k)  \nonumber\\
		&\,   \oplus \mathbf   L \cdot \textbf{Enc}( Q(\bar y(k)   \,\,\text{mod}\,\,q  ) \\
		\!\!\!\!\! \mathbf{\tilde v} (k+1) \!&= \mathbf S \cdot   \mathbf{	\tilde v } (k)   \nonumber\\
		&\, \oplus\! \frac{s }{\gamma} \! \cdot  \! \textbf{Enc}\left(\! Q\left(\! \frac{S}{s} \tilde v_p(k) - \frac{S}{s} \tilde v(k)\right)\text{mod}\,q \!\right) \\
		\!\!\!\!\! \mathbf{\bar	u}(k) \! &= \mathbf  K \cdot \mathbf{\tilde x}(k) \oplus \mathbf{(V-K \Gamma)} \cdot  \mathbf {\tilde  v}(k)  	 
	\end{align}
\end{subequations}
in which $\mathbf{\tilde x}(k) \in \mathcal C ^{n}$, $\mathbf{\bar u}(k) \in \mathcal C ^{w}$ and $\mathbf{\tilde v}(k) \in \mathcal C ^{w}$ are the cihpertexts of $\tilde x(k)$, $\bar u(k)$ and $\tilde v(k)$, respectively. Its initial conditions are given by \textbf{Enc}$(\tilde x(0)\,\text{mod}\, q)$ and \textbf{Enc}$(\tilde v(0)\,\text{mod}\, q)$. The matrices in (\ref{controller 5}) follow
\begin{align*}
	\begin{array}{l}
		\mathbf{A-LC}:=  \frac{A-LC}{\gamma} \,\text{mod}\,q ,   	\mathbf{B}:=  \frac{s  B}{ \gamma} \,\text{mod}\,q , 
		\mathbf{L}:=  \frac{s  L}{\gamma} \,\text{mod}\,q , \\
		\mathbf{K}:=  \frac{K}{s } \,\text{mod}\,q ,  
		\mathbf{V-K\Gamma}:=  \frac{V-K \Gamma}{s }  \,\text{mod}\,q  ,   	\mathbf{S}:= \frac{S}{\gamma} \,\text{mod}\,q
	\end{array}
	%	& \mathbf{A-LC}:=  \frac{A-LC}{\gamma} \,\text{mod}\,q , 
	%	\mathbf{B}:=  \frac{s  B}{ \gamma} \,\text{mod}\,q , \\
	%	&\mathbf{L}:=  \frac{s  L}{\gamma} \,\text{mod}\,q , 
	%	\mathbf{K}:=  \frac{K}{s } \,\text{mod}\,q , \\
	%	&\mathbf{V-K\Gamma}:=  \frac{V-K \Gamma}{s }  \,\text{mod}\,q  ,  
	%	\mathbf{S}:= \frac{S}{\gamma} \,\text{mod}\,q.
\end{align*}
where $q$ is the modulus of the cryptosystem and will be specified later. 
Note that $\mathbf  {\bar u}(k)$ in (\ref{controller 5}a) is generated by the encrypted controller (\ref{controller 5}c) instead of being generated by the actuator through the re-encryption technique\cite{kimtac1}. 

\textbf{Algorithm on the actuator side}:
We implement the following algorithm
\begin{align}
	&\!\!	\left\{\!\!
	\begin{array}{l}\label{u actuator}
		\bar u_a(k) \!= \! \mathbf{Dec(\bar u}(k)) \!-\! \left\lfloor \! \frac{ \mathbf{Dec(\bar u}(k))  + C_ v \bar U_a(k-1)+ \frac{q}{2}}{q} \! \right\rfloor q \\
		u_a(k)=l(k) \bar u_a(k)
	\end{array}\right.\\
	&\!\! \left\{\!\!
	\begin{array}{l}\label{u actuator 2}
		%C_v=	[c_{v-1}\, c_{v-2}\, \cdots c_0]\\
		\bar 		U_a(k-1)\!:=\! [\bar u_a^T(k\!-\!1)\,\,\bar u_a^T(k\!-\!2)\cdots\bar u_a^T(k\!-\!v) ]^T \\
		%	\bar u_a(k-j)= \frac{u_a(k-j)}{l(k-j)},\quad j=1, 2, \cdots, v\\
		C_ v:=	[c_{v-1}\, c_{v-2}\, \cdots c_0] \in \mathbb Z^{1\times v}
	\end{array}\right. 
\end{align}
in which $c_{v-1},\, c_{v-2},\, \cdots, c_0$ are the coefficients in the characteristic polynomial of $\bar S: = S/\gamma \in \mathbb Z ^{v\times v}$:
\begin{align}\label{poly}
	\text{det}(\lambda I_v- \bar S)= \lambda^v + c_{v-1} \lambda^{v-1}+ \cdots + c_0 I_v.
\end{align}
To compute $u_a(k)$, the actuator should have a memory to store the previous $\bar u_a(k-j)$ with $j=1, 2, \cdots, v$. 
By (\ref{controller 5}) and (\ref{u actuator}), one can see that the controller only transmits $\mathbf{\bar u} (k)$ to the actuator at $k$.

In the following lemma, we show that the control input can be represented by the previous control inputs leveraging the Cayley–Hamilton theorem.

\begin{lemma}\label{lemma 2}
	%\begin{align} 
	%	&\! \! \! \bar u(k)= -c_{v-1} \bar u(k -1) - c_{v-2} \bar u (k -2)\cdots -c_0 \bar u(k-v)\nonumber\\
	%	&\! \!  \! + \! a(k) \!+\! c_{v-1} a(k\!-\!1)   \! + \!c_{v-2} a (k\!-\!2)\! \cdots + \! c_0 a(k\!-\!v).
	%\end{align}
	The control input $\bar u(k)$ in (\ref{controller 4}c) is equivalent to
	\begin{align} \label{bar u}
		\bar u(k)= - C_ v \bar U (k-1) + C_ {v+1} Z(k)
	\end{align}
	with $C_v$ in (\ref{u actuator 2}) and
	\begin{subequations}
		\begin{align}\label{a(k)}
			&C_{v+1}:=[1\,\, c_{v-1}\, c_{v-2}\cdots c_0] \in \mathbb Z ^{1\times (v+1)}\\
			& \bar	U(k-1)\!:=\! [\bar u^T(k\!-\!1)\,\,\bar u^T(k\!-\!2)\cdots\bar u^T(k\!-\!v) ]^T \\
			&	Z(k):= [z^T(k)\,\, z^T(k-1)\,\, \cdots\,\, z^T(k-v)]^T\\
			&	z(k):= K \bar p(k) - K \bar e_x (k) - (V- K \Gamma)\bar e_v (k)\\
			&\bar e_x(k): = \bar x_p(k) - \bar x(k),\quad \bar e_v(k): = \bar v_p(k) - \bar v(k)\\
			&\bar p(k): = \bar x_p(k) - \Gamma \bar v_p(k).
		\end{align}
	\end{subequations}
\end{lemma}

\textbf{Cayley–Hamilton theorem \cite{horn2012matrix}:} Before presenting the proof of Lemma \ref{lemma 2}, we introduce the Cayley–Hamilton theorem as follows. For any matrix $F\in \mathbb R ^{n\times n}$, $F^n$ satisfies
\begin{align}\label{27}
	F^n = - c_{n-1} F^{n-1}   - c_{n-2} F^{n-2}-\cdots-c_0I_n
\end{align}
in which $c_{n-1}\cdots c_0$ follow those in the characteristic polynomial of $F$ as det$(\lambda I_n-F)= \lambda^n + c_{n-1} \lambda^{n-1}+ \cdots + c_0 I_n$.  

\textbf{Proof.}
By the definitions of $\bar p(k)$, $\bar e_x (k)$ and $\bar e_v (k)$, $\bar u(k)$ in (\ref{controller 4}c) is equivalent to the following form
\begin{align}\label{20}
	\bar u(k)  \! =\! V \bar v_p (k) \!+ \! K \bar p(k) \!- \! K \bar e_x (k) \!-\! (V\!-\!K\Gamma) \bar e_v (k).     
\end{align}
Then the dynamics of $\bar v_p(k)$ and $\bar u(k)$ can be written as
\begin{subequations}\label{controller dynamics}
	\begin{align}
		\bar v_p(k+1)&=\bar S \bar v_p(k) \\
		\bar u(k)&= V \bar v_p(k) + z(k)
	\end{align}
\end{subequations}
in which $\bar S= S/\gamma \in \mathbb{Z}^{v\times v}$.
By the Cayley–Hamilton theorem, one has 
$
\bar S^v = - c_{v-1} \bar S^{v-1}   - c_{v-2} \bar S^{v-2}-\cdots-c_0I_v.
$
Therefore, by $\bar v_p(k)= \bar S ^v \bar v_p(k-v)$,
one can obtain
\begin{align*}
	\bar v_p (k) =( - c_{v-1} \bar S ^{v-1} - c_{v-2} \bar S ^{v-2}\cdots - c_0 I_n ) \bar v_p(k-v).
\end{align*}
By (\ref{controller dynamics}b), we have 
\begin{align}
	\bar u(k)&=-V  \sum_{j=1}^{v} c_{v-j}\bar S ^{v-j} \bar v_p(k-v) + z(k)\nonumber\\
	&=-V  \sum_{j=1}^{v}c_{v-j} \bar v_p(k-j) + z(k)\nonumber\\
	&= -\sum_{j=1}^{v}c_{v-j} [\bar u(k-j) - z(k-j)] + z(k)
\end{align}
which implies the result in (\ref{bar u}).\qedp

If $u_a(k)=u(k)$ for all $k$, we say that the actuator is able to restore the control input. 
We are ready to present the main result of the paper.

\begin{theorem}\label{Theorem 1}
	Consider the encrypted controller (\ref{controller 5}) and the algorithm (\ref{u actuator}) operated in the actuator. If the modulus
	\begin{align}\label{proposition 1 q}
		q> 2\| C_{\bar S} ^{v+1}\|_\infty \left(2\|K\|_\infty C_{p,e}+
		\frac{ \|V-K\Gamma\|_\infty}{2\gamma}\right)
	\end{align}
	then one has $u_a (k)= u(k)$ for all $k$ with $u(k)$ in (\ref{controller 2}c). In (\ref{proposition 1 q}), $C_{p,e}$ is given in the Appendix. Moreover, the asymptotic tracking control problem in (\ref{tracking}) is solved.
\end{theorem}
\textbf{Proof.}
We conduct the proof by induction. We mainly show that if the actuator is able to restore the previous control inputs such that $ \bar U_a(k-1)=  \bar U(k-1)$, then it can also restore $u(k)$ by obtaining $\bar u_a(k)=\bar u(k)$.

Note that one can only obtain the lower bits of $\bar u(k)$ by decryption in light of $ \mathbf{Dec(\bar u}(k)) =  \bar u(k)\,\text{mod}\, q $. Then, (\ref{u actuator}) is equivalent to
\begin{align} \label{proposition 2 34}
	\bar u_a(k)	&=   \bar u(k)\,\text{mod}\, q  \!-\! \left\lfloor\! \frac{ \bar u(k)\,\text{mod}\, q  + C_ v  \bar U_a(k-1)+\frac{q}{2}}{q}\!\right\rfloor \! q \nonumber\\
	& =  \bar u(k)  - \left\lfloor \frac{ \bar u(k)    +C_ v  \bar U_a(k-1)+\frac{q}{2}}{q}\right\rfloor q.  
\end{align}				
We are interested if $\| \bar u(k)    +C_  v  \bar U_a(k-1)\|_\infty$ is upper bounded by $\frac{q}{2}$. In light of $\bar u(k)$ in (\ref{20}), $C_ v$ in (\ref{u actuator 2}) and $ \bar U_a(k-1)=  \bar U(k-1)$ by hypothesis, one has 
\begin{align}\label{35 new}
	\bar u(k)    + C_ v  \bar U_a(k-1)
	%	& =  a(k) + c_{v-1} a(k-1)   + c_{v-2} a (k-2)\cdots +  c_0 a(k-v) \nonumber\\
	=  C_ {v+1}Z(k).
\end{align}				
Therefore, one has 
\begin{align}\label{norm}
	\!\!\!\! \|\bar u(k)   \! + \! C_ v  \bar U_a(k-1)\|_\infty   \le \|C_ {v+1}\|_\infty \|Z(k)\|_\infty \!<\! \frac{q}{2}.  
\end{align}	
To derive the inequality above, we have applied the following inequalities $\| Z(k)\|_\infty\le \|[K\,\,-K] \|_\infty \| [\bar p^T(k)\,\, \bar e_x ^T(k)]^T\|_\infty + \|V+K\Gamma\|_\infty\|\bar e_v(k)\|_\infty $, in which $ \|[\bar p^T(k)\,\, \bar e_x ^T(k)]^T\|_\infty  \le  \|[\bar p^T(k)\,\, \bar e_x ^T(k)]^T\| \le C_{p,e}$ and $\|\bar e_v(k)\|_\infty    \le \frac{1}{2\gamma}$  will be shown in the Appendix. 
Under (\ref{norm}), one should have $\left\lfloor \frac{ \bar u(k)    +C_ v  \bar U_a(k-1)+q/2}{q}\right\rfloor=0$, and therefore $\bar u_a(k)= \bar u(k)$ in view of (\ref{proposition 2 34}). Then, it is simple to obtain $u_a(k)= l(k)\bar u_a(k)=l(k) \bar u(k)=u(k)$.

To show asymptotic tracking control, it is sufficient to show 
$
\| y(k) - v_p(k) \|_\infty =\| Cx_p(k) - C\Gamma v_p(k)\|_\infty=  l(k) \|C\bar p(k) \|_\infty \le l(k) \|C\|_\infty C_{p,e} 	\to 0
$, in which $\|\bar p(k)\|_\infty \le C_{p,e}$ and $l(k) \to 0$ as $k \to \infty$. \qedp

\begin{remark}\label{remark 2}
	In Theorem \ref{Theorem 1}, we are able to find a finite $q$ to restore $u(k)$ on the actuator side. 
	%In particular, we have shown that there exists a finite $q$ and provided a lower bound of it. 
	It is worth mentioning that $C_ v \bar U_a(k-1)$ in (\ref{u actuator}) is the key establishment of ensuring a finite $q$. 
	It is simple to verify that $\bar u(k)=  u(k) / l(k) \to \infty$ as $l(k)\to 0$. 
	Then, for any finite $q$, if one removes $C_v  \bar U_a(k-1)$, one must encounter $\lfloor \frac{ \bar u(k)+ q/2}{q} \rfloor \ne 0$ after some $k$. Then, it is not possible to obtain $\bar u_a(k)=\bar u(k)$ by (\ref{u actuator}) or equivalently (\ref{proposition 2 34}). Thus, we implement $C_v \bar U_a(k-1)$ to ``counteract" the growth of $\bar u(k)$ such that $\bar u(k)    +C_v  \bar U_a(k-1) $ does not diverge, and hence we are able to find a finite $q$.  
	% such that $\tilde x(k)  - (\frac{A}{\gamma}\tilde x(k-1) + \frac{sB}{\gamma}\bar u(k-1))$ is bounded by $\frac{q}{2}$.
	% Eventually, one is able to recover $\tilde x(k)$ by (\ref{17}). 
	We mention that if $q$ is allowed to be infinite, e.g., $q> 2\|\bar u(k)\|_\infty$, one can simply implement $\mathbf{Dec(\bar u}(k)) - \left\lfloor  \frac{ \mathbf{Dec(\bar u}(k))  + q/2}{q}  \right\rfloor q$ to restore $\bar u(k)$. 
\end{remark}

\begin{remark}\label{remark 3} 
	One can write $\bar u(k) = \bar u_1 (k)q + \bar u(k) \,\text{mod}\, q$ for some $u_1(k) \in \mathbb{Z}^w$.  
	Recall the definitions of ``higher bits" and ``lower bits" in the Notation. One can see that the actuator actually receives only the lower bits of $\bar u(k)$. If we follow the methods in \cite{farokhi2017secure,cheon2018need}, in which the controller transmits both the higher and lower bits to the controller, $q$ should cover all the possible $\bar u(k)$. However, due to $\bar u(k)\to\infty$, covering all the possible $\bar u(k)$ by a finite $q$ is not possible. After some $k$, the higher bits of $\bar u(k)$ will be lost during the decryption process $ \mathbf{Dec(\bar u}(k)) =  \bar u(k)\,\text{mod}\, q $. Under such a situation, our method can still enable the actuator to restore $\bar u(k)$ under a finite $q$.
	\qedp 
\end{remark}

%\begin{remark}\label{remark 4} 
%	Our algorithm (\ref{u actuator}) provides the possibility of detecting if $\bar u(k)$ belongs to the space of plaintexts $\mathbb Z_q$ or not. 
%	%The algorithm  is able to detect the overflow problem presented . 
%	For a finite $q$, 
%	as $\bar u(k)$ grows and exceeds $q/2$, $\mathbf{Dec}(\mathbf{ \bar u} (k))= \bar u(k)\,\,\text{mod}\,\,q$ will lose the information of $\bar u(k)$. Therefore, the detection of overflow is difficult \cite{cheon2018need}.  
%	%First, we present why it is not possible to detect overflow by only utilizing $\mathbf{Dec}(\mathbf{ \bar u} (k))= \bar u(k)\, \text{mod}\, q$. 
%	%If $\|\bar u(k)\|_\infty < \frac{q}{2}$, then $\mathbf{Dec}(\mathbf{ \bar u} (k))= \bar u(k)\, \text{mod}\, q=\bar u(k)$. As $\bar u(k)$ grows and exceeds $\frac{q}{2}$, 
%	%the decrypted value $ \bar u(k)\, \text{mod}\, q$ is not able to reflect overflow. However, it is clear that when overflow occurs, $\mathbf{Dec}(\mathbf{ \bar u} (k))= \bar u(k)\, \text{mod}\, q\ne\bar u(k)$. By contrast, the actuator is able to detect the occurrence of $\|\bar u(k)\|_\infty \ge \frac{q}{2}$.
%	In (\ref{norm}), we have shown that $ \|\bar u(k)   \! + \! C_{\bar S} ^v U_v(k-1)\|_\infty  < \frac{q}{2}$, which implies $ -\frac{q}{2} +C_{\bar S} ^v U_v(k-1)< \bar u(k) < \frac{q}{2} + C_{\bar S} ^v U_v(k-1)$. Since $\tilde x(k)$ is known by the actuator, it is able to know when $\bar u(k)$ exceeds $q/2$. We have also shown that when $\bar u(k)$ exceeds $q/2$, the actuator can still restore $\bar u(k)$ from $\mathbf{Dec}(\mathbf{ \bar u} (k))$.  \qedp
%\end{remark}

\subsection{Unbounded internal state and the solution}

\textbf{Unbounded internal state:} In Section 3.1, though the algorithm on the actuator side is able to restore the control input in light of $u_a(k)=u(k)$, one would encounter the problem of unbounded internal state. That is, the internal state $\bar u_a(k)$ is unbounded: $\|\bar u_a(k)\|_\infty=\|u(k)/l(k)\|_\infty \to \infty$ in the actuator as $\lim_{k \to \infty}l(k)=0$. The issue of unbounded internal state also exists in \cite{kimcdc}, see ``\textbf{Dec}($\mathbf{u}(t)$ mod $\left(  q, u(t-1)/(s_1s_2l(t))     \right)$" in (31) in \cite{kimcdc}.

In the field of systems and control, in generally one should design a control system whose state is bounded. Moreover, in practice, numerical saturation in electronic devices would impact an unbounded state. That is, if saturation occurs, $\bar u_a(k)$ in (\ref{u actuator}) is upper bounded and therefore one has $\bar u_a(k)\ne \bar u(k)$. This implies that the actuator must fail to restore the correct control input after some $k$ due to $u_a (k)= l(k) \bar u_a(k) \ne l(k) \bar u(k) = u(k)$. 
In this subsection, we propose a control scheme whose internal state is bounded.

First, we present the fundamental idea by plaintexts. We assume that the controller and the actuator have memory units to store previous control inputs in ciphertexts and plaintexts, respectively. At $k$, the controller transmits
\begin{align}\label{29}
	m(k):=\bar u(k)+C_v   \bar U(k-1)
\end{align}
to the actuator, in which $\bar u(k)$ follows that in (\ref{bar u}). The actuator stores 
previous ``control inputs"
$U_a(k\!-\!1):= [  u_a ^T(k\!-\!1)\,\,  u_a ^T(k-2)\cdots  u_a ^T(k-v) ]^T
$ in the memory. If the actuator is able to restore previous control inputs, namely, $U_a(k-1)= U(k-1):= [  u ^T(k\!-\!1)\,\,  u^T(k-2)\cdots  u ^T(k-v) ]^T$, then it can also restore $u(k)$ in view of
\begin{align}\label{46}
	&l(k)m(k)-C_v  \text{diag}(\gamma,\gamma^2, \cdots, \gamma^v)U_a(k-1)  \nonumber\\
	&= l(k)(\bar u(k)  +C_v \bar U(k-1))-l(k)C_v \bar U(k-1)  \nonumber\\
	& =u(k)
\end{align} 
in which 
$
C_v  \text{diag}(\gamma,\gamma^2, \cdots, \gamma^v)U_a(k-1)=l(k)C_v \bar U(k-1)
$. Importantly, 
$m(k)=\bar u(k)+C_v   \bar U(k-1)= C_  {v+1} Z(k)$ is bounded and hence there must exist a finite $q$ such that
\begin{align}\label{mk}  
	m(k)\,\text{mod}\, q  - \left\lfloor \frac{m(k)\,\text{mod}\, q    +\frac{q}{2}}{q}\right\rfloor q  
	=  m(k), \forall k. 
\end{align}		
Overall, by (\ref{29})--(\ref{mk}), one can see that the actuator should be able to restore $u(k)$ by utilizing $m(k)$, which in particular is a bounded state. In the following encrypted control scheme, we will use $m(k)$ as the internal state to restore $u(k)$ instead of $\bar u_a(k)$ in Section 3.1.

\begin{figure}[t]
	\begin{center}
		\includegraphics[width=0.48\textwidth]{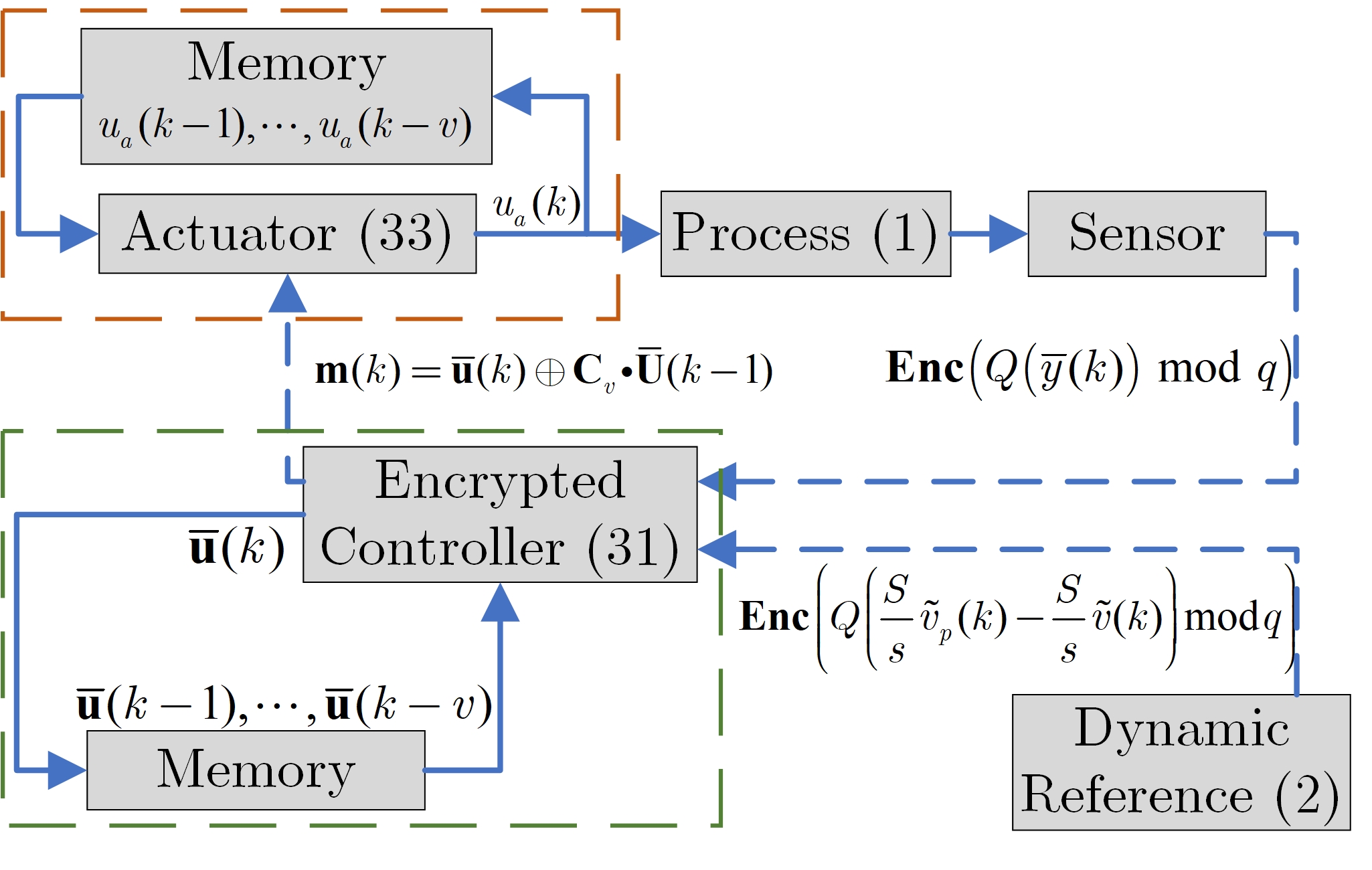}  \\
		\caption{Encrypted control architecture. Dashed lines represent networks. The sensor and reference provider send the same messages as in Fig. \ref{Control archi 1}. The encrypted controller generates $\mathbf{\bar	u}(k)$ and $\mathbf{m}(k)$. It transmits $\mathbf{m}(k)$ to the actuator over networks and stores $\mathbf{\bar	u}(k)$ in the memory for being utilized at $k+1$. The actuator computes $m_a(k)$, and $u_a(k)$ based on $u_a(k-j)$ ($j= 1, \cdots, v$). $u_a(k)$ is then fed to the process and stored in the memory for being utilized at $k+1$.} \label{Control archi 2}
	\end{center}
\end{figure}

In Fig. \ref{Control archi 2}, we present the encrypted control architecture. The encrypted controller and the algorithm on the actuator side are provides as follows.

\textbf{Encrypted controller:}
Based on the idea of the control scheme above, we present the controller over encrypted data:
\begin{align}\label{51}
	\left\{
	\begin{array}{l}
		(\ref{controller 5}a)-(\ref{controller 5}b)\\
		\mathbf{\bar	u}(k)  = \mathbf  K \cdot \mathbf{\tilde x}(k) \oplus \mathbf{(V-K \Gamma)} \cdot  \mathbf {\tilde  v}(k) \\ 	 
		\mathbf m(k):=\mathbf{\bar u}(k) \oplus \mathbf C _ v \cdot \mathbf{\bar U}(k-1)
	\end{array}
	\right.
\end{align}
in which $\mathbf C _ v:=C_ v \,\,\text{mod}\,\,q$, $\mathbf m(k) \in \mathcal C ^ w$ is the output of the controller,  and 
\begin{align}
	\mathbf{\bar U}(k-1):= [ \mathbf{\bar u} ^T(k\!-\!1)\,\,  \mathbf{\bar u}^T(k-2)\cdots \mathbf{\bar u}^T(k-v) ]^T
\end{align}
is available at $k$ thanks to the memory unit in the controller. Note that $C_v$ has all integer elements because $\bar S$ is an integer matrix, and hence its characteristic polynomial has only integer coefficients. (\ref{51}) can be further simplified into a system consisting of (\ref{controller 5}) and $\mathbf m(k)$. However, for highlighting that $\mathbf{\bar	u}(k)$ is a necessary state for computing $\mathbf m(k)$ and should be also stored in the memory, we do not simplify (\ref{51}).

\textbf{Algorithm on the actuator side}:
When the actuator receives $\mathbf{m}(k)$ from the controller, it calculates
\begin{subequations}\label{52}
	\begin{align}
		& m_a(k):= \textbf{Dec}(\mathbf{m}(k)  )  - \left\lfloor \frac{\textbf{Dec}(\mathbf{m}(k)  )   +\frac{q}{2}}{q}\right\rfloor q  \\
		&  u_a(k)  = l(k)m_a(k)   - C_v  \text{diag}(\gamma,\gamma^2, \cdots, \gamma^v)U_a(k - 1). 
	\end{align}
\end{subequations}

\begin{proposition}\label{Proposition 1}
	Consider the encrypted controller (\ref{51}) and the algorithm on the actuator side (\ref{52}). If (\ref{proposition 1 q}) holds, then one has $m_a(k)=m(k)$ and $u_a(k)=u(k)$. Moreover, the tracking control problem in (\ref{tracking}) is solved.
\end{proposition}

\textbf{Proof.} We conduct the proof by induction. If $U^a(k-1)= U(k-1)$, then one should obtain $u_a(k)= u(k)$.
First, note that
\begin{align*}
	m_a(k) &=	m(k)  - \left\lfloor \frac{	m(k) +\frac{q}{2}}{q}\right\rfloor q  \\
	=&	m(k)   - \left\lfloor \frac{	C_{\bar S} ^{v+1} Z(k)    +\frac{q}{2}}{q}\right\rfloor q =m(k)
\end{align*}
where $\left\lfloor \frac{	C_{\bar S} ^{v+1} Z(k)    +\frac{q}{2}}{q}\right\rfloor q=0$ because of $\|	C_{\bar S} ^{v+1} Z(k)   \|_\infty < \frac{q}{2}$ (see (\ref{norm})). 
Substituting $m_a(k)=m(k)$ into (\ref{52}b) and then following (\ref{46}), one can obtain $u_a(k)=u(k)$. \qedp

\begin{remark}
	We compare the control schemes in Sections 3.1 and 3.2. First, note that they require the same modulus $q$ in (\ref{proposition 1 q}). Second, the actuator in Section 3.1 has $\bar u_a(k)$ as an internal state, which becomes infinitely large as $k\to \infty$. Whereas the actuator in Section 3.2 has $m_a(k)$ as the internal state, which is upper bounded in light of $\|m_a(k)\|_\infty=\|m(k)\|_\infty=\|C_{v+1}Z(k)\|_\infty<q/2$. Third, only the actuator in Section 3.1 needs to store previous control inputs in the memory. Whereas in Section 3.2, the controller and the actuator should store the ciphertexts and plaintexts of previous control inputs, respectively.    \qedp
\end{remark}
%Compared with the previous algorithm on the actuator side (\ref{u actuator}), the new algorithm (\ref{52}) does not require the actuator to compute $U(k-1)\to \infty$.  

\color{black}

\section{Simulation}
In this section, we conduct simulation to verify the results of this paper. The matrices of the process ($A$, $B$, $C$), the reference dynamic matrix ($S$),  the feedback gain ($K$) and observer gain ($L$) are given and calculated as follows
\begin{align}
	A=
	\begin{bmatrix}
		0 &0\\
		0& 0.5
	\end{bmatrix}, 
	B=
	\begin{bmatrix}
		1 &-1\\
		0& 2
	\end{bmatrix}, 
	C=
	\begin{bmatrix}
		0.1 &1\\
		0& 0.1
	\end{bmatrix}, \nonumber\\
	S=
	\begin{bmatrix}
		1.5 &2\\
		0& 1
	\end{bmatrix}, 
	K=
	\begin{bmatrix}
		0 &-0.5\\
		0& 0
	\end{bmatrix}, 
	L=
	\begin{bmatrix}
		0 &5\\
		0& 0
	\end{bmatrix}. \nonumber 
\end{align}
We select $\gamma=0.5$ and $s=0.5$, under which all the matrices in (\ref{controller 4}) contain only integers.

\begin{figure}[t]
	\begin{center}
		\includegraphics[width=0.4 \textwidth]{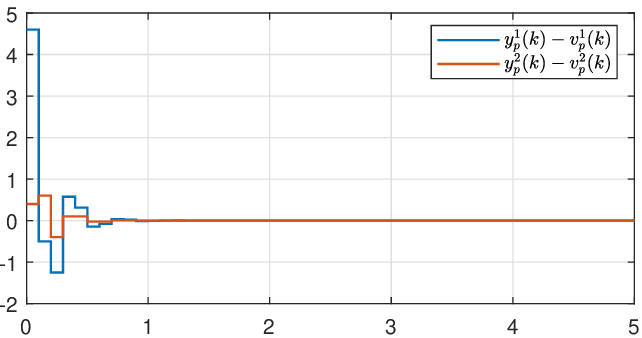} \\
		\linespread{1}\caption{Time responses of tracking errors $y_p(k)-v_p(k)$ } \label{Fig3}
	\end{center}
\end{figure}

\begin{figure}[t]
	\begin{center}
		\includegraphics[width=0.41\textwidth]{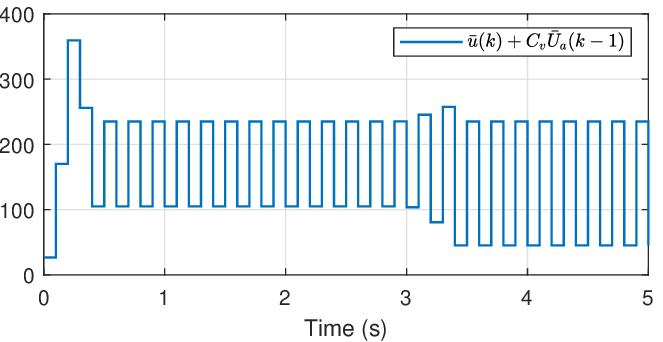}  \\
		\caption{Time responses of $\| \bar u(k)    +C_v \bar U_a(k-1)\|_\infty$. } \label{Fig4}
	\end{center}
\end{figure}

We first show the simulation results corresponding to Theorem \ref{Theorem 1}. Fig. \ref{Fig3} presents the time responses of tracking errors, in which one can see that the tracking errors converge to zero. In Theorem \ref{Theorem 1}, one of the key tasks is to ensure that $\| \bar u(k)    +C_ v  \bar U_a(k-1)\|_\infty$ is finite and therefore one is able to find a finite $q$. 
As shown in Fig. \ref{Fig4}, $\| \bar u(k)    +C_ v\bar U_a (k-1)\|_\infty$ does not exceed 359.5 in the simulation horizon. According to the simulation result, selecting $q=2^{10}>2\times359.5$ is sufficient. Meanwhile, according to the theoretical result in Theorem \ref{Theorem 1}, $q$ should be larger than 16878, which implies that one should select $q=2^{15}$. The conservativeness between the theoretical result and the simulation result is because we have followed a ``worst case" type of analysis, in which have frequently used ``$\le$", ``$\max$" and ``$\|Dx+Ey\| \le \|D\|\|x\| + \|E\|\|y\|$" (for some matrices $D$ and $E$, and some vectors $x$ and $y$). Because Proposition \ref{Proposition 1} restores the same $u(k)$ and requires the same $q$, the simulation results of Proposition \ref{Proposition 1} will be very similar to those in Figs. \ref{Fig3} and \ref{Fig4}, and hence are omitted.

\section{Conclusions}

This paper investigated asymptotic tracking control of dynamic reference over homomorphically encrypted data with a finite modulus.
%We found overflow problems caused by dynamic reference and asymptotic tracking, though the controller coefficients are integers, which prevents the actuator from restoring the correct control input. 
We designed a tracking controller with only integer coefficients leveraging the zooming-in factor of dynamic quantization, under which the re-encryption technique is not required. 
Exploiting the Cayley-Hamilton theorem, we represented the control input as a linear combination of previous control inputs. Therefore, the algorithm on the actuator side is able to restore the control inputs with a finite modulus from the lower bits. A lower bound of the modulus is also provided in the paper. Secondly, we solved the problem of unbounded internal state in the actuator, by formulating a new controller output and algorithm on the actuator side. The actuator can restore the correct control input under the same modulus as in the first result.

\appendix

\section*{Appendix}
We will show that there exists a finite $C_{p,e}$ satisfying $\|[	\bar  p^T(k)\,\,
\bar 	e_x ^T (k) ]^T\| < C_{p,e}$.
We first present the dynamics of $\bar p(k)$, $\bar e_x(k)$ and $\bar e_v(k)$:
\begin{subequations}\label{25}
	\begin{align}
		\begin{bmatrix}
			\bar p(k+1) \\
			\bar e_x(k+1)
		\end{bmatrix}& =
		\underbrace{\frac{1}{\gamma}\begin{bmatrix}
				A+BK & -BK\\
				0 &A-LC
		\end{bmatrix}}_{:=A_{cl}}
		\begin{bmatrix}
			\bar p(k) \\
			\bar e_x(k)
		\end{bmatrix} \nonumber\\ 
		& \quad+ 
		\underbrace{\begin{bmatrix}
				\frac{-B(V-K\Gamma)}{\gamma} & 0 \\
				0 & \frac{L}{\gamma}
		\end{bmatrix}}_{:=B_{cl}}
		\begin{bmatrix}
			\bar e_v(k) \\
			\bar e^q _y(k) 
		\end{bmatrix} \\
		%	&\bar p(k+1) = \frac{A+BK}{\gamma} p(k) - \frac{BK}{\gamma} \bar e_x(k) \nonumber\\
		%	&\quad\quad \quad\quad  - \frac{B(V-K \Gamma)}{\gamma} \bar e_v (k)\\
		%	&\bar e_x (k+1) = \frac{A-LC}{\gamma} \bar e_x (k) + \frac{L}{\gamma} \bar e ^q _y(k) \\
		\bar e_v(k+1) &= \frac{S \bar e_v(k)}{\gamma} - \frac{1}{\gamma} Q(S \bar e_v(k)). 
	\end{align}
\end{subequations}
in which $\|\bar e_v(k)\|_\infty =\frac{1}{\gamma} \| S \bar e_v(k) -  Q(S \bar e_v(k))\|_\infty \le \frac{1}{2\gamma}$ and $ 
\| \bar e^q _y(k)\|_\infty : = \|\bar y(k) - Q(\bar y(k))\|_\infty \le \frac{1}{2} $.   
By the result of $\gamma$ in Lemma \ref{lemma 1}, it is clear that $A_{cl}$ is a Schur matrix. Then there exist $0<\rho<1$ and $C_\rho$ such that 
$
\left\|
A_{cl} ^k
\right\| \le C_\rho \rho ^k.  
$
Hence, by (\ref{25}a), one can obtain
\begin{align}\label{28}
	&\left\|	[
	\bar p^T(k+1)\,\, 
	\bar e_x ^T (k+1)
	]^T
	\right\|
	\le 
	C_\rho \rho ^{k+1}
	\left\|	[
	\bar p^T(0)\,\, 
	\bar e_x ^T (0)
	]^T
	\right\| \nonumber\\ 
	&\quad\quad\quad\quad\quad\quad\quad   + C_\rho \sum_{i=0}^{k} \rho^{k-i}
	\left\|B_{cl} \right\|
	\left\|	\begin{bmatrix}
		\bar e_v(i) \\
		\bar e^q _y(i) 
	\end{bmatrix}
	\right\|
\end{align}
%in which $	\|\bar e_v(i)\|_\infty  \le \frac{1}{2\gamma},\| \bar e^q _y(i)\|_\infty  \le \frac{1}{2} $ as mentioned earlier after (\ref{25}).
in which the following inequalities hold:
\begin{align}
&C _ \rho \rho ^{k+1}
\|	[
\bar p ^T (0)\,\,
\bar e_x ^T (0)
]^T
\| \nonumber\\
& \le C_\rho \rho ^{k+1} \sqrt{2n}
(\|x_p(0)\|_\infty + \|\Gamma\|_\infty \|v_p(0)\|_\infty)/l(0) \\
&C _\rho \sum_{i=0}^{k}  \rho^{k-i} 
\|	B_{cl} \|
\|	[\bar e_v ^T (i) \,\,
{\bar e _y}^{qT } (i) 
]^T	\|  \nonumber\\
&\le
C _\rho \left\|B_{cl} \right\|\frac{\sqrt{2v}}{2\gamma(1-\rho)} (1-\rho^{k+1}).
\end{align}
% ($n$ and $v$ are the dimensions of $x_p(k)$ and $v_p(k)$, respectively)
Recalling $\|x_p(0)\|_\infty \le C_{x_p(0)}$ and $\|v_p(0)\|_\infty \le C_{v_p(0)}$,	one can calculate $C_{p,e}$ as 
\begin{align}\label{31}
	& 	\left\|	[
	\bar p^T(k+1)\,\, 
	\bar e_x ^T (k+1)
	]^T
	\right\| \nonumber\\
	& \le 
	\max\left\{ C_\rho \frac{\sqrt{2n}(C_{x_p(0)} + \|\Gamma\|_\infty C_{v_p(0)})}{l(0)},\right. \nonumber\\ 
	&\left.  \quad\quad\quad \quad   C_\rho \left\|	
	B_{cl}
	\right\|\frac{\sqrt{2v}}{2\gamma(1-\rho)}   \right\} =:C_{p,e}.
\end{align}				

\bibliographystyle{unsrt}
\bibliography{bib}

\end{document}